\documentclass[12pt,english,aps,manuscript]{revtex4}
\usepackage{graphicx}
\usepackage{amssymb,amsfonts,amsmath,verbatim}
\usepackage{epstopdf}
\usepackage{setspace}

\begin{document}
\preprint{Scientific Reports}
\title{The anatomy of urban social networks and its implications in the searchability problem}% Force line breaks with \\

\author{C Herrera-Yag\"ue$^{1,2,3}$, CM Schneider$^1$, T Couronn\'e$^4$, Z 
    Smoreda$^4$, RM Benito$^{1,5}$, PJ Zufiria$^{2,3}$ and MC 
    Gonz\'alez$^{1}$\footnote{E-mail: martag@mit.edu}
}
\address{$^1$Department of Civil and Environmental Engineering, Massachusetts
Institute of Technology, Cambridge, MA, USA}
\address{$^2$Depto. Matem\'atica Aplicada a las Tecnolog\'ias de la 
    Informaci\'on y las Comunicaciones (TIC),
ETSI Telecomunicaci\'on, Universidad Polit\'ecnica de Madrid (UPM), Spain}
\address{$^3$C\'atedra Orange. Universidad Polit\'ecnica de Madrid (UPM), Spain}
\address{$^4$Sociology and Economics of Networks and Services department, 
    Orange Labs, Issy les Moulineaux, France}
\address{$^5$Grupo de Sistemas Complejos, Departamento de F\'isica y 
    Mec\'anica, ETSI Agr\'onomos, Universidad Polit\'ecnica de Madrid (UPM), 
    Spain}

\begin{abstract}
The appearance of large geolocated communication datasets has recently 
increased our understanding of how social networks relate to their 
physical space. However, many recurrently reported properties, such as 
the spatial clustering of network communities, have not yet been 
systematically tested at different scales. In this work we analyze the 
social network structure of over 25 million phone users from three 
countries at three different scales: country, provinces and cities. We 
consistently find that this last urban scenario presents significant 
differences to common knowledge about social networks. First, the 
emergence of a giant component in the network seems to be controlled by 
whether or not the network spans over the entire urban border, almost 
independently of the population or geographic extension of the city.  
Second, urban communities are much less geographically clustered than 
expected. These two findings shed new light on the widely-studied 
searchability in self-organized networks. By exhaustive simulation of 
decentralized search strategies we conclude that urban networks are 
searchable not through geographical proximity as their country-wide 
counterparts, but through an homophily-driven community structure.  
\end{abstract}
\maketitle

\section*{Introduction}
In the last decade social network analysis methods have allowed us to uncover 
local and global patterns~\cite{leskovec2008planetary}, locate influential individuals 
\cite{kitsak2010identification}, and examine network dynamics~\cite{rybski2009scaling}.  The study 
of macro-level social networks traces the outcomes of collective and 
large-scale social interactions such as economic development~\cite{eagle2010network}, 
resource transfer~\cite{ugander2012structural}, disease transmission~\cite{watts2005multiscale}, 
and communications~\cite{onnela2007structure} over a large population. In these 
cases,  networks nodes represent individuals, and links are generally defined 
by friendships or acquaintances among them. Well documented structural patterns 
of these networks are: the positive correlations in the degree of adjacent 
nodes~\cite{newman2003social}, a short diameter (increasing as the natural 
logarithm of the number of nodes)~\cite{watts1998small}, and network 
transitivity or clustering, which is the propensity for nodes pairs to be 
connected if they share a mutual neighbor~\cite{watts1998small}.  
Interestingly, social networks are also divided in groups or communities, and 
the existence of such communities alone can produce  both degree correlations 
and high clustering~\cite{newman2003structure}.  On the other hand, some social 
links are also the consequence of similar attributes of their nodes. Similar 
people tend to select each other~\cite{centola2011experimental,christakis2007spread}, they communicate 
more frequently and present stronger social 
interactions~\cite{onnela2007structure}. 

Parallel to the rise of social network analysis, and often using 
similar data sources, human mobility patterns have also considerably 
evolved in the recent years \cite{gonzalez2008understanding, 
    de2013unique}. An interesting topic of study which has started to 
grow recently is  to combine findings from both areas to explain the 
relationship between social networks and geographical space. Evidently, 
social contacts can exist only if there is the opportunity for such 
contacts to be created.  This explains, for example,  the ubiquitous 
findings showing that geographic proximity  favors the existence of 
social contacts~\cite{liben2005geographic,noulas2012tale}.  
Additionally, Network Communities (locally dense areas of the social 
graph) have been analyzed in several country scale social networks when 
the spatial positions of the nodes are known, as more and more often is 
the case in social networks resulting from information and 
communication technologies \cite{expert2011uncovering}. A 
well-documented result of these communities is that they retrace 
national \cite{blondel2010regions, thiemann2010structure} and 
administrative borders \cite{ratti2010redrawing} when studied at a 
country scale.

The spatial dispersion of social contacts at country scale has been 
studied in the context of transportation planning (see 
\cite{kowald2013distance} and references therein).  Kowald et al. made 
a comparative study of social ties and their distances, from surveyed 
of individuals within cities in three different continents.  They 
reported that although the ultimate models need to incorporate the 
characteristics of egos, ties, and transportation facilities, there is 
a general trend of a power law decay of social ties with distance. In 
this work we want to explore the group structure social networks in 
cities and its relation to space.  The size of social groups have 
important implications our societies, Simmel \cite{simmel1971} viewed 
the increasing size in networks groups as the origin of the isolation 
of individuals.  These implications and the related literature are out 
of the scope of this work.  Social networks studies within cities have 
measured the role of the density of social ties 
\cite{bettencourt2007growth,pan2013urban} or face to face encounters 
\cite{calabrese2011interplay, sun2013understanding}.  Here, we are 
interested in the analysis of communities within cities and their 
relation to space.  Despite some analysis of communities within cities 
\cite{walsh2011spatial,gao2013discovering}, there is still lack of 
knowledge on a clear structure of urban social networks in space.  
Specifically, how connected components emerge with distance 
\cite{barthelemy2011spatial} in urban social networks. 

%Within cities the possibilities of face-face interactions are favored 
%\cite{calabrese2011interplay} due to the nature of urban organization 
%\cite{bettencourt2007growth,pan2013urban}, but not much is known about 
%community structure of social networks at the urban scale compared to the 
%observations at the national scale. Few studies have analyzed network's 
%communities within cities and their relation to space 
%\cite{walsh2011spatial,gao2013discovering}, but we still lack a conclusive 
%picture of the structure of urban social networks. Finally, while some 
%theoretical work has been done characterizing the phase transitions associated 
%with the emergence of a giant component in spatial networks 
%\cite{barthelemy2011spatial}, empirical evidence on whether or not people 
%living in a certain area form a connected social network has not been 
%presented to date.

Here we assign each mobile phone user to a fixed location corresponding 
to his/her most commonly used zipcode or mobile phone tower with the 
goal of systematically studying the spatial properties of their social 
networks at different scales, including the formation of a giant 
component in space. The geographic distance between two nodes is then 
defined as the distance of their respective most common locations, 
typically home or work. It is expected that within cities this distance 
should not be a strong limiting factor in the creation of social ties 
as it may be other factors that define their social distance. Social 
distance is given by differences between groups of society, including 
differences such as socio-demographic, race or social identity  
\cite{bogardus1947measurement}. Searchability is a well-established 
network property that relates to both geographic proximity and social 
distance: ordinary people are capable of directing messages only 
through their acquaintances and to reach any target person in only a 
few steps. Milgram \cite{milgram1967small, travers1969experimental} 
first discovered this property, in a
social experiment that routed letters across U.S.  In the light of email 
communication, Dodds et al. \cite{dodds2003experimental} showed that 
when routing a message to a target, people selected in the first steps 
acquaintances that could be geographically close to that target.  
However, in the latests steps, participants selected acquaintances that 
could belong to the professional group of the target (i.e.  socially 
close).  Up to now, the network structure that makes searchability 
possible has not been empirically measured in large-scale social 
networks.

 We designed our study to explore the role of both social and geographic
 distances in social networks. Social distance is not 
 trivially defined in social networks with data passively collected without 
 much information about the attributes of the nodes. Introducing a metric of 
 social distance for these cases is an interesting question, but out of the 
 scope of this work.  Watts et al. defined the social distance between 
 two nodes as the difference in hierarchy levels of the two smallest 
 groups the nodes belonged to \cite{watts2002identity}. Here, we use a 
 similar definition, proposed by Kleinberg et al.  
 \cite{kleinberg2006complex}: social distance $S(u,v)$ between nodes 
 $u$ and $v$ is the number of nodes in the smallest group containing 
 $u$ and $v$. In this work we define social groups as network 
 communities, which are locally dense sub-networks.  Networks 
 communities are thus a central aspect to the analysis of social 
 networks, being the source of their structural properties (degree 
 correlation and high clustering) and consequence of non-structural 
 properties, such as homophily \cite{mcpherson2001birds}.  The 
 detection of network communities (modules or groups) is a difficult 
 task that has attracted much attention in the last few years 
 \cite{fortunato2010community}.  Here we adopt a well-established 
 method that detects communities by optimizing the Newman Girvan 
 modularity metric \cite{newman2004finding}. 

We first present a general description of the measured social networks, with
focus on the small-world properties and link-distance distributions. Next, we
report the performance of different routing strategies and show that geogreedy
strategies (choosing the smallest geographical distance to the target) 
are
ineffective within cities while strategies based on social distance (choosing
within the smallest community) still work. We discover two features of urban
social networks that cause the failure of geographic strategies: urban
communities are geographically dispersed and there is not a large connected
component in groups of nodes defined by their geographic proximity. We further
measure in the urban networks how the density of links $P(u,v)$ decays with
increasing group size ($S$) or distance. We find that the probability of finding
a link between individuals $u$ and $v$ in a group of size $S$ scales as $P(u,v) 
\sim
S^{-\gamma}$, with $\gamma < 1$ when groups ($S=S_r$) are defined by users 
living within geographic
balls of a certain radius $r$.  This is in contrast with observations at the 
national scale which
report $\gamma >= 1$ \cite{liben2005geographic}. These results support the evidence that while
geogreedy algorithms work to reach a target's city, they fail within urban
borders. In addition, we show that the condition $\gamma >= 1$ still holds when
groups ($S=S_c$ )  are defined by social distance. These results of urban groups
defined by either social distance or geographic distance are in nice agreement
with the analytic conditions of networks searchability 
\cite{kleinberg2000navigation} and support the
results reported in routing experiments \cite{dodds2003experimental}. This work provides novel evidence
of social networks: urban networks form geographically dispersed 
communities
that make them searchable. 

\section*{Results}
\subsection*{Network Structure}
\begin{figure}[htp]
   \centerline{
   \includegraphics[width=1\textwidth]{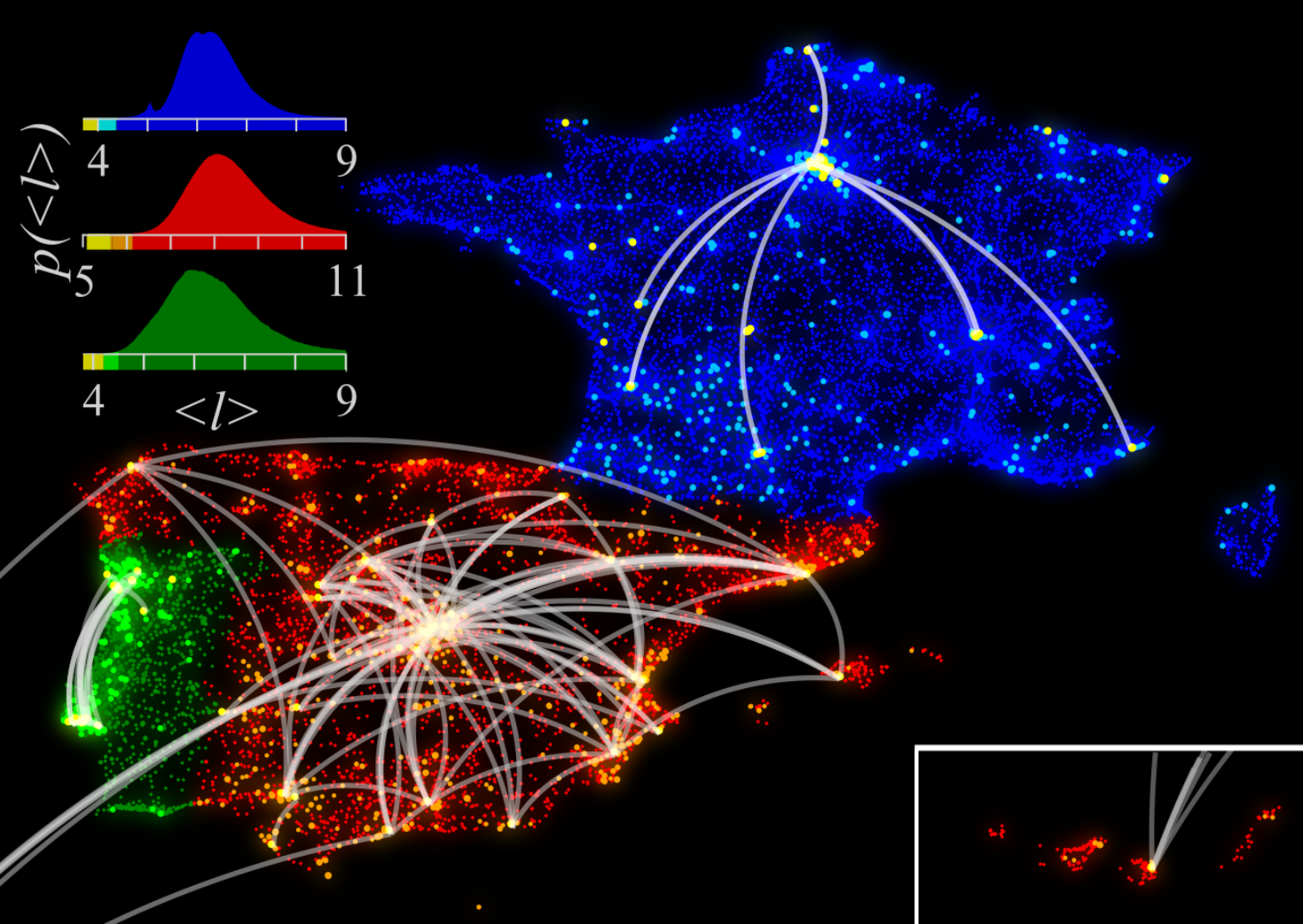}}
   \caption{Visualization of central places in France, Spain and Portugal. Each 
       circle represents a mobile phone tower and its color (the brighter the 
       more central) corresponds to the inverse of closeness centrality 
       $\langle l \rangle$ (average number of hops to any other person) 
       of the most central people in this tower. People are always 
       assigned either to their billing address or most used tower.  
       White lines highlight the social network between the 50 most 
       central persons of each country. In the three insets the 
       distribution of the  $\langle l \rangle$ of all persons and the 
       relation to the used color are also shown. This figure was 
       created using Grace and Inkscape.}
   \label{Fig1}
\end{figure}

Our data set contains information for 7 billion mobile phone interactions 
gathered during a 6 month period in France, Portugal and Spain. We report the 
structural network characteristics in table \ref{Table:param}.
These results confirm that the networks exhibit the \emph{small world} 
property, with the average number of people in the shortest path between a 
sender and a recipient $\langle l \rangle$ is $6.5$, $6.4$, and $8.4$ in the 
different countries, similar to the values reported in previous 
works~\cite{onnela2007structure,eagle2009inferring}. As a sole illustration of 
the resulting networks, we extract the spatial distribution of the most 
central people in the network, considering someone is more central if 
he/she is in average closer to everyone else in the graph (closeness 
centrality).  In Fig.  \ref{Fig1} we show the distribution of the 
average graph distance between a sender and all possible recipients 
$p(\langle l \rangle)$ among the population for each country. This 
value is also known as the inverse of the closeness 
centrality~\cite{freeman1977set} and it ranges from $3.8$ to $11$, so 
everyone in the country is in average within $4$ hops from the most 
central people and within $11$ of the less central ones.  Each dot 
represents a mobile phone tower, which is our smallest spatial 
resolution.  In order to expose the backbone of the social network, the 
color intensity of each mobile phone tower represents the closeness 
centrality of the most central person in that tower.  Additionally, the 
links highlight the social connections only among the $50$ most central 
people in each country, showing significant differences in the social 
network analyzed in the three countries 

%While main cities appear brighter, centrality is not only determined by 
%population density: Barcelona area (Spain NE) is highly populated but it seems 
%to be socially less central than Alicante (SE) even though the latter has half 
%the population. Notice also, that the most geographically central city is not 
%necessarily the most socially central one.  . Whereas in France, Paris hosts 
%the most central people, in Portugal two cities, Lisbon and Porto, seem to be 
%equally important. In contrast, the most central people in Spain are spread 
%over the entire country including even the Gran Canaria and Mallorca islands.  
%Current spatial network models would not reproduce these differences observed 
%in the data, but they would rather place central actors in the geographical 
%center of the territory and/or the most populated cities. This suggests that 
%the relationship between social networks and their underlying geography is yet 
%to be fully understood and further research is needed in this topic. 

\begin{figure}[h]
   \centerline{
   \includegraphics[width=1\textwidth]{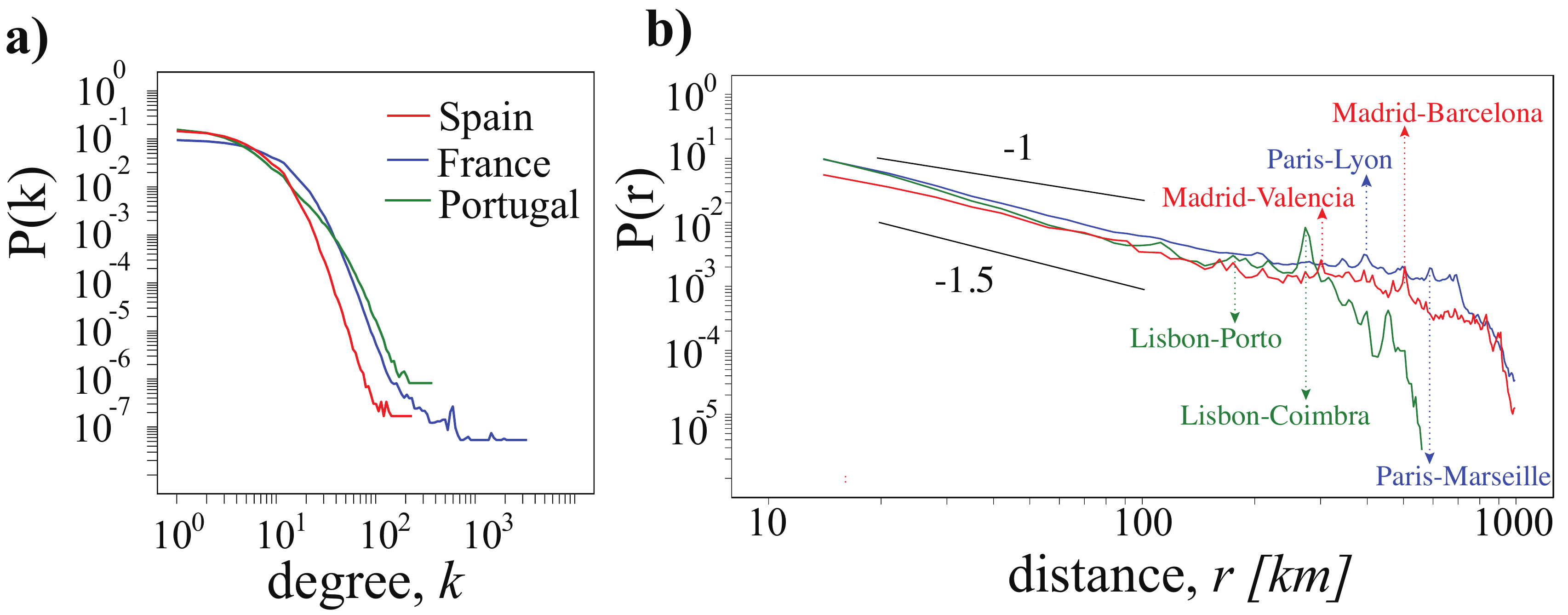}}
   \caption{Country-wide social networks structure. (a) Degree distribution for 
       each of the country level networks. (b) Probability of a link to have 
       distance $r$ in each of the networks. Distances are grouped in 7 km 
       bins.  In all three countries, distribution present a power law decay 
       (exponents between $-1$ and $-1.5$) up to $100$ km. A large fraction of 
       links lie within the same tower ($r$ = 0), averaging $40 \%$ in Spain 
       (red), $18\%$ in France (blue) and $21\%$ in Portugal (green). }
   \label{Fig2}
\end{figure}

Regarding degree distribution, our three networks present the common heavy-tail 
distribution found in previous works with social 
networks~\cite{onnela2007structure, lambiotte2008geographical}. Degree 
distributions for all three networks are shown in Fig.  \ref{Fig2}a (details 
about power-law fitting can be found in Table S1).  We note the existence of 
hubs (nodes with very high number of connections) in all three networks.  In 
order to measure geographic proximity  between individuals we need to assign a 
location to each of them. In our study, users are located in their billing zip 
code (Spain) or their most used tower (France and Portugal).  Spain zip codes 
are geolocated according to geonames database, available at 
http://downloads.geonames.org/export/zip, and grouped according to latitude and 
longitude since some zip codes have identical coordinates. Towers coordinates 
were provided by the carrier.  In total $8,928$ different locations are 
available in Spain, $17,475$ in France and $2,209$ in Portugal.  It is well 
documented that the probability of finding a social tie decreases with 
geographic proximity, regardless the proxy used to infer the social network: 
blogs \cite{liben2005geographic}, location based social networks 
\cite{scellato2011exploiting, cho2011friendship} or mobile phone data 
\cite{lambiotte2008geographical,onnela2007structure,blondel2010regions}. In all 
of them the fraction of social links between nodes that are within distance $r$ 
from each other decreases (at least in a certain range) as a power law, with 
exponents between $-1$ and $-2$.  As shown in Fig.  \ref{Fig2}b, our 
data fits this behavior for all three networks. Kowald et al.  
\cite{kowald2013distance} present a careful analysis of the decaying 
function observing distance bands depending on the population, similar 
analysis on this data remain to further studies.

Moreover, due to the high number of links considered we are able to 
observe long-range peaks.  The reason for these peaks is the 
heterogeneity in the spatial distribution of population (we observe the 
same peaks even if we randomize the links while keeping actors in the 
same location).  Once established that the short paths exist all across 
the network, we explore the success of routing strategies at two 
levels: intercity and intracity.

\subsection*{Exploring Routing Strategies}

In order to gather insights on the social network structure, we investigate the 
well-known searchability condition. We explore different routing strategies on 
the social networks described above. We separate the routing experiment 
into two phases: intercity routing and intracity routing.

Intercity routing seeks to reach the correct city while intracity 
routing searches for the individual target within a city.  Cities are 
defined by their administrative borders.  In this study we consider two 
scales: provinces and municipalities as shown in  Fig. S5.  On both 
phases, we test different decentralized routing strategies which employ 
only information of neighbor nodes (also called contacts or friends).  
In a random search ({\bf ran}), individuals route the message by 
randomly selecting a neighbor node that did not have had the message 
previously.  Geographical routing ({\bf geo}) passes the message to the 
contact that is geographically closest to the final target, whereas 
degree routing ({\bf deg}) selects the friend with the highest number 
of friends.  Finally, community routing ({\bf com}) forwards the 
message to a friend such that he/she belongs to the smallest community 
containing the target (see details in the {\bf Methods} section). 

\begin{figure}[h]
   \centerline{
   \includegraphics[width=0.9\textwidth]{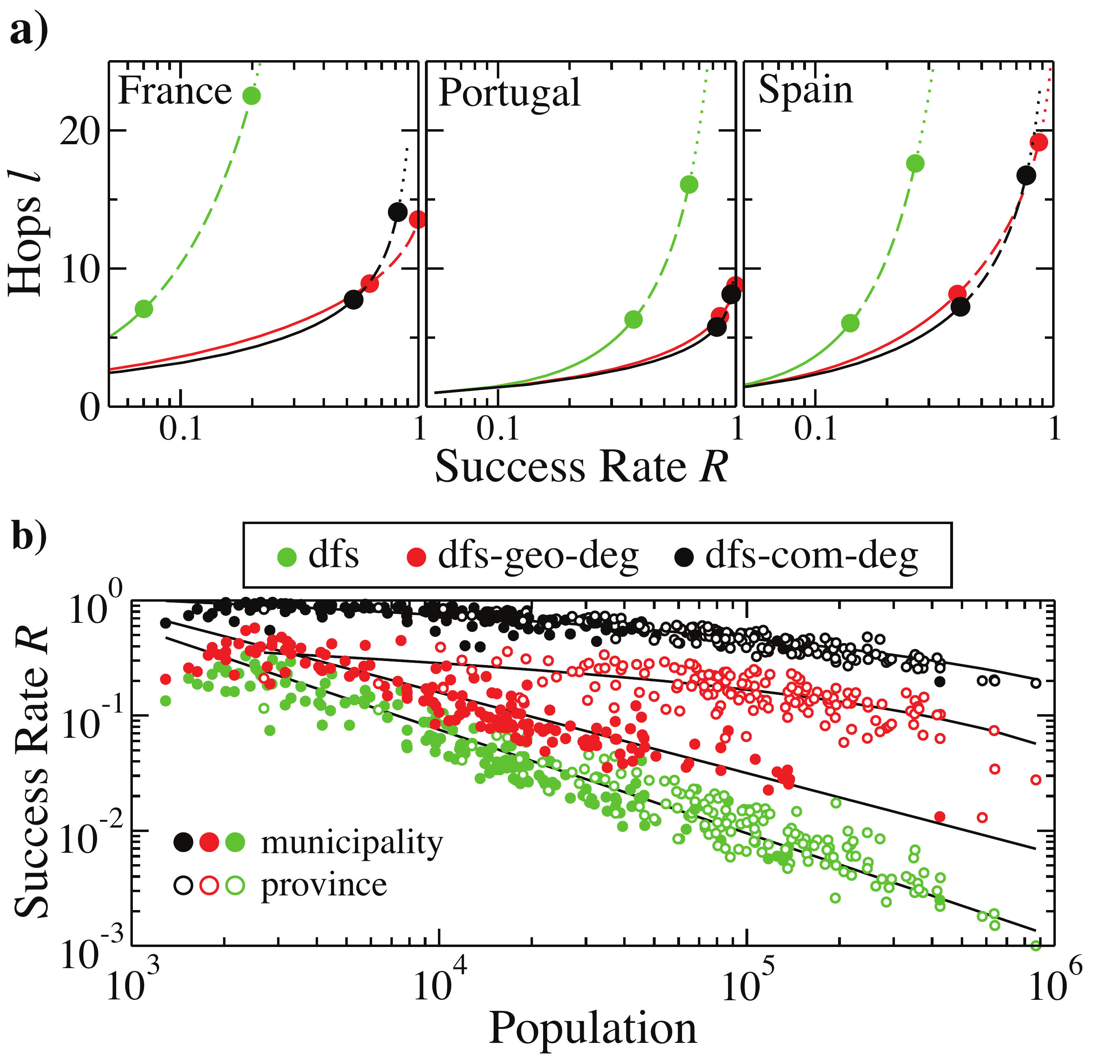}}
   \caption{Results for different routing strategies in both stages. a) 
       Dependence of the number of hops l on the success rate $R$ for intercity 
       routing (results for completing the delivery within $15$ and $100$ hops 
       are highlighted by circles). b) Success rate versus population 
       size for three strategies in 155 municipalities and 150 
       provinces.  All logarithmic and power-law functions are guides 
       to the eye.}
    \label{Fig3}
\end{figure}

 Our intercity simulations results presented in Fig. \ref{Fig3}a indicate that 
 both \emph{geo} and \emph{com} routing are able to reach the target cities.  
 Moreover, the success rate depends only logarithmically on the population size 
 of the destination city (Fig. S8), confirming that both strategies are equally 
 efficient. The intercity experiment can be replicated in our 
 homepage~\footnote{Herrera-Yag\"{u}e, C. Finding Bacon.  
     http://humnetlab.mit.edu/findingbacon/ Accessed 2014 Nov 1 (2014).}.  
 Geographic strategies had already been reported successful using a half 
 million bloggers network across the US~\cite{liben2005geographic}.  However, 
 intracity routing has not been previously explored because both the low sample 
 size of the network ($0.15\%$ of US population) and the lack of information of 
 the coordinates of individuals within cities obliged to relax the modeled 
 network structure: namely, nodes were allowed to forward messages to anyone 
 else within the target city, even if they were not directly connected. In 
 contrast, our larger population sample ($12\%$ - $40\%$) and much smaller 
 spatial resolution (mobile phone tower scale) allow us to explore routing 
 inside cities using strict routing among connected individuals. 

Next, we explore routing strategies by analyzing the network properties 
within the geographic administrative borders at two scales: provinces 
as upper limits (usually containing large cities plus suburbs) and 
municipalities as lower limits (see SI for details). Thus, we analyze 
the three different routing strategies in $155$ social networks from 
the large municipalities and all $150$ provinces of the three 
countries. In contrast to intercity routing, routing inside 
municipalities is significantly more successful if the strategy uses 
community information (Figs. S10 - S15 show additional strategies). For 
different routing strategies Fig. \ref{Fig3}b shows the success rate 
for municipalities (filled circles) and provinces (open circles) in 
each country as a function of the population size N; an upper limit of 
100 hops was employed and Fig. S24 shows results with a smaller upper 
limit. We find that at both scales the community based routing is 
efficient because of the slow decay in success rate $R \sim c - b \ln 
N$ ($c = 2 \pm 0.03$ and $b = 0.133 \pm 0.003$) and in contrast to the 
random strategy, which as expected decays almost reversely linear as  
$R \sim N^{-a}$  ($a = 0.95 \pm 0.03$). Interestingly, the 
geographically based routing presents a crossover behavior between 
municipalities (only intracity routing) and provinces (including an 
initial intercity stage). This behavior is due to the fact that a 
province consists of several municipalities. Although the 
geographically based routing reaches the correct municipality, within 
the municipality this strategy fails. This explains the different 
scaling observed for geographic routing in municipalities and 
provinces: while within municipalities the routing success rate scales 
similarly to the random routing $R \sim N^{-a}$  ($a = 0.66 \pm 0.03$), 
the province routing success rate scales similarly to community routing 
$R \sim c - b \ln N$ ($c = 0.82 \pm 0.05$ and $b = 0.056 \pm 0.004$), 
but with a lower success rate as a consequence of its inefficacy within 
municipalities.

In the next sections we show that the failure of the geographic routing within 
cities lies in two previously unknown spatial properties of urban social 
networks: lack of short-range connectivity and geographical dispersion of urban 
communities.

\subsection*{Connectivity collapse within cities}

\begin{figure}[h]
   \centerline{\includegraphics[width=0.85\textwidth]{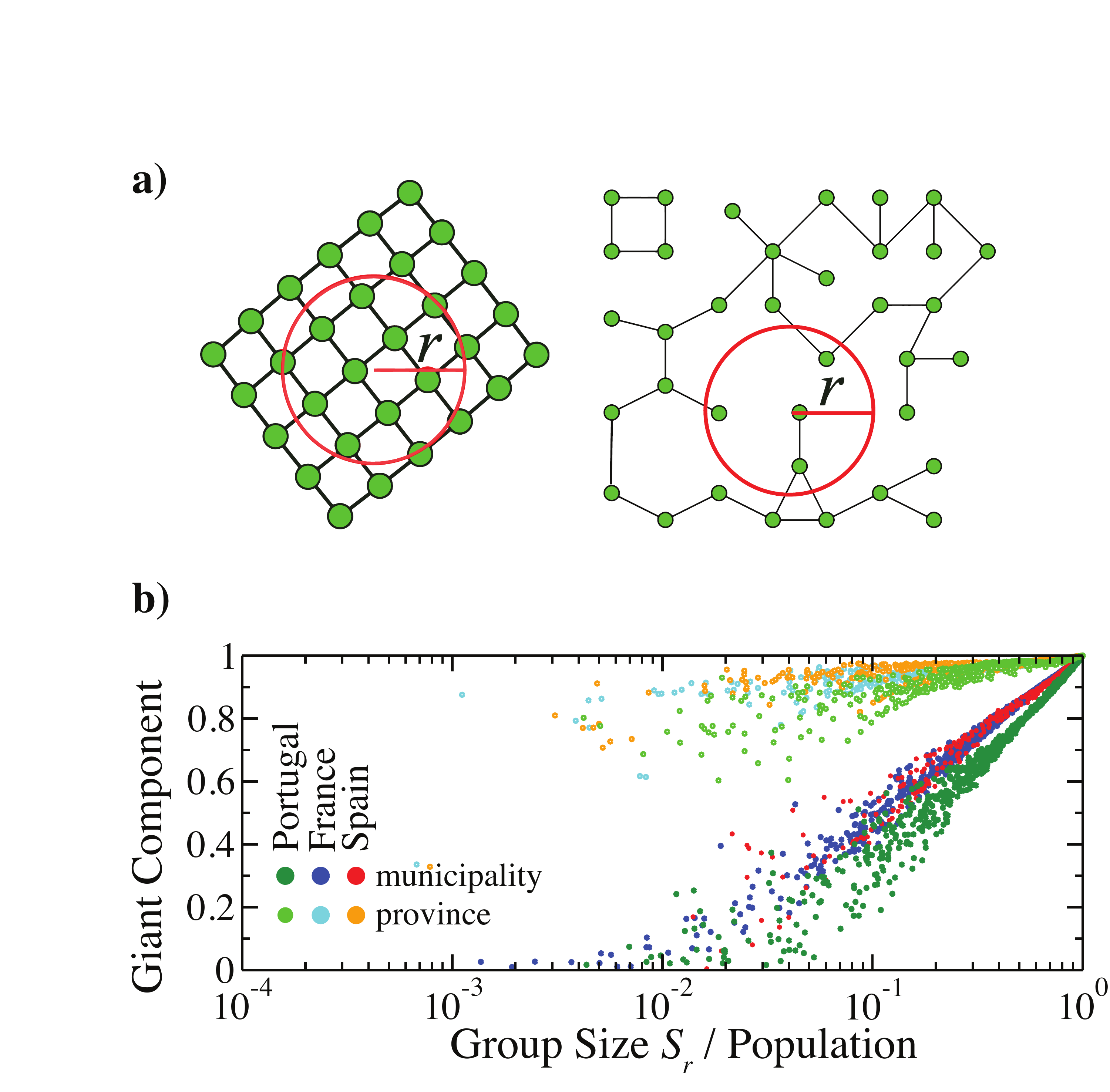}}
   \caption{
   Short range connectivity
   a) In a 2D lattice (left), any geographic ball contains a connected network, 
   however this is not the case for any network (right) where the path between 
   two nodes within a geographic ball might include nodes out of the ball if 
   the network induced by the nodes within the ball is not connected.
   b) Fraction of nodes in the giant component as a function of the 
   relative size of the geographic ball for the three capitals compared 
   to the country-wide networks.  Each of the $6000$ dots in the figure 
   was calculated by selecting 2 nodes $u$ and $v$ at random within a 
   city or within the country, extracting the subnetwork defined by the 
   ball whose center is in $u$ and radius up to $v$, and identifying 
   the number individuals that belonged to the giant component of such 
   subnetwork.}
    \label{Fig4}
\end{figure}

A necessary condition for any geogreedy algorithm to succeed in a routing 
experiment is that the subgraph induced by the nodes located within any 
geographic ball of radius $r$ must be connected. This is equivalent to saying 
that if a message headed to target user B has reached a user A, A and B 
in the same connected component within the subgraph induced by those 
nodes included in the circle whose center is in B and has radius up to 
A.  While this is granted in a lattice our results show that is not 
necessarily the case in a real-world network (see Fig.  \ref{Fig4}a).  
We test this structure in our data using geometric and social 
distances. We
divide the network into groups of size $S_{X}$ using either geographic 
balls (while in this work we only consider 2D geographic \emph{circles} 
we keep the term \emph{balls} for consistency with previous theoretical 
work \cite{kleinberg2006complex} which has been generalized to higher 
dimensions) of
a certain radius $r$ ($X = r$) or existing communities ($X = c$)
A natural question emerging then is: which is the critical radius $r_{c}$ so 
that geographic balls with $r>r_{c}$ are likely to contain a connected network?  
Interestingly we observe that there is not a unique $r_{c}$, but rather this 
radius is defined by the size of a city, so that only geographic balls 
containing entire cities contain a connected network.

We illustrate this fact further by calculating the size of the largest 
connected component within different radius  and group sizes, performing this 
analysis centered in different locations from the capital municipality (city) 
or centered in a province of the three countries. Fig. \ref{Fig4}b shows that 
the fraction of nodes in the giant component is much smaller within cities than 
within provinces. Surprisingly, we find that this lack of connectivity is not 
caused by not having enough short-distance links (actually between $18\%$ and 
$40\%$ of links are within the same location (tower or zip-code)). When we zoom 
into a region of the city we find small highly clustered groups  which form 
islands; the paths among these geographically neighboring groups exist through 
people living far away. 

\begin{figure}[h]
   \centerline{\includegraphics[width=0.95\textwidth]{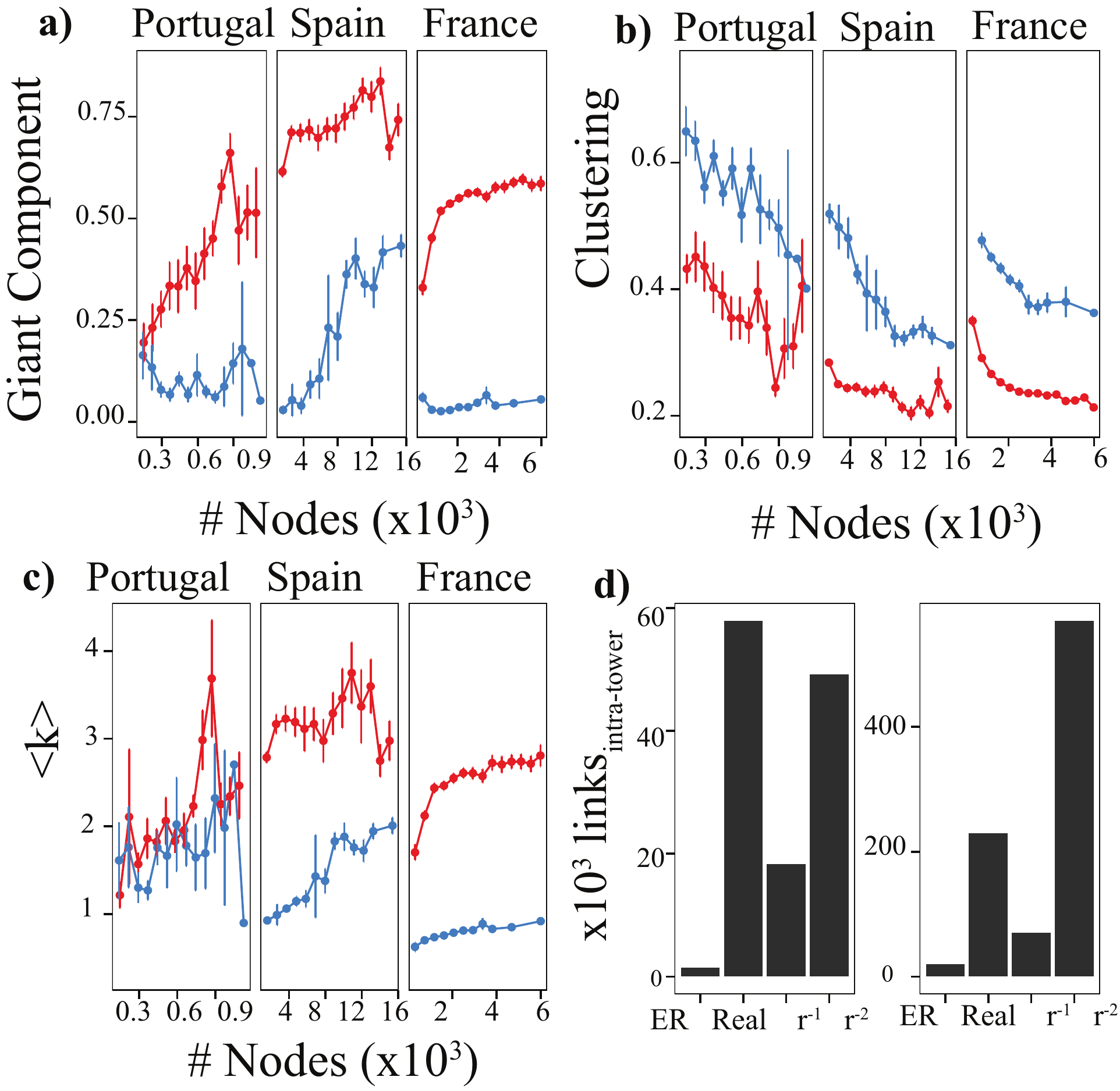}}
   \caption{Connectivity collapse within cities. a) Relation between population size and fraction of nodes in the giant component for all towers in the capital cities (blue) and municipalities in the country within the same range of population (red). Errors bars represent the standard
error of the mean $\frac{\sigma}{\sqrt{n}}$. The size of the connected 
components within municipalities tends to be higher than within towers 
of the same size. b) and c) depict the causes of this behavior, smaller 
average degree and higher clustering are the reasons why the giant 
component is larger in municipalities. d) Number of links within the 
same tower using several randomization models. Results are averaged 
over $10$ runs.
The real network has a bigger number of intra-tower links than a space 
independent graph (ER) and a $\frac{1}{r}$ model. In the case of Lisbon, the 
real network has even more links than a $\frac{1}{r^{2}}$ model. To explain the 
high number of intra-tower links the geographical distance is not sufficient, 
thus another effect like clustering is needed.}
   \label{Fig5}
   \end{figure}

To better illustrate this finding we have studied all intra-tower networks in 
the capital cities and compared them to networks of the same size centered in 
municipalities in the countryside. Fig. \ref{Fig5}a shows the average giant 
component for towers and municipalities of a certain size.  Municipalities with 
a given population have a larger giant component than a tower in a city with 
the same population.

Given a fixed number of nodes, a giant component emerges more likely with a 
higher number of links and with low clustering (a link closing a triangle does 
not enlarge any connected component). As shown in Figs. \ref{Fig5}b and 
\ref{Fig5}c, both effects are present at the municipality level and not within 
towers. This explains the different giant component sizes between 
municipalities and towers.  However, high clustering seems to be dominant for 
the lack of a connected component, since in Portugal the average degree is the 
same in towers and municipalities.  Moreover, the small average degree does not 
seem to be due to lack of data, since the data from France presents the highest 
average degree at a country scale, while it exhibits the smallest average 
degree on the tower scale.

Our results on geographic distance $r$ agree with previous literature 
\cite{lambiotte2008geographical,liben2005geographic} showing that the 
probability of two users within distance $r$ to be connected follows $P(r) \sim 
\frac{1}{r}$. However, this sole finding does not give us any information about 
the number of links between people within the same location (tower/zipcocode), 
since in principle they are within $r=0$ distance. In order to be able to apply 
pure geographical models (generating links with $P(r) \sim 
\frac{1}{r^{\alpha}}$) to our data, we have to randomize the position of the 
users around the tower's location. A common assumption for mobile phone data is 
considering that if a call is processed by a tower, then that tower is the 
closest to the user's location.  This assumption implies that the geographic 
space can be divided according to the Voronoi diagram of the towers in that 
region.  This way our randomization assigns each user a position uniformly 
distributed in the Voronoi cell it belongs to. Figure S22 shows the 
randomization process in Paris and Lisbon. After randomization, the distance 
$r$ between any two users is greater than zero, so we can apply 
$\frac{1}{r^{\alpha}}$ models the number of predicted and present intra-tower 
links for the same number of links in the whole network. In Fig.  
\ref{Fig5}d we show that the number of observed intra-tower links in 
both cities is higher than what a pure geographical model $1/r$ would 
generate (even higher than a $1/r^2$ in the case of Lisbon). Despite 
this abundance of links, there is no giant component, what implies that 
clustering plays a major effect at this level, producing highly 
clustered \emph{islands }within the same tower. 

\subsection*{Geographical dispersion of urban communities}

\begin{figure}[h]
\noindent \begin{centering}
\includegraphics[height=0.7\textheight]{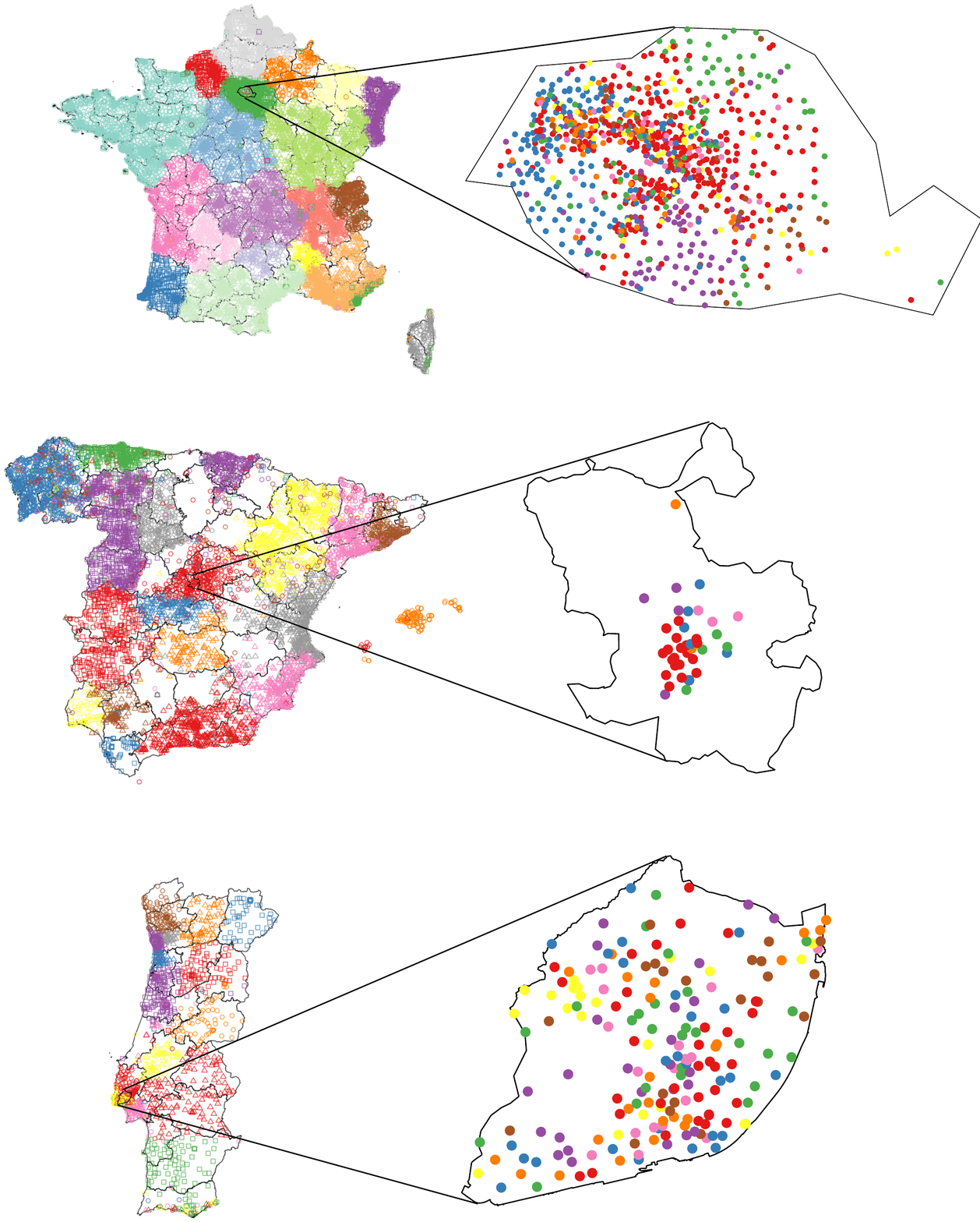} \par\end{centering}

\caption{\label{fig:Geographical-clustering-of}Geographical clustering of
social communities. On the country scale, towers belonging to the
20 biggest communities are presented in different colors and shapes.
On the city scale, towers within each capital city are presented.
On the country scale most of communities fit with the administrative
boundaries while within cities communities do not seem to be 
geographically driven. The figure was created using R packages 
\emph{maptools} and \emph{ggplot2}.}
    \label{Fig6}
\end{figure}

On the country scale the identified communities are known to be highly 
spatially correlated and even redraw the administrative borders as shown in 
Fig. \ref{Fig6} (left) where the colors indicate the dominant community of each 
mobile phone tower.This has been the motivation of a research line oriented
to \emph{redraw} the political maps according to social network features
\cite{blondel2010regions,ratti2010redrawing,calabrese2011connected}.
However, in the city scale (Fig.  
\ref{Fig6} right) the communities are dispersed over space and within the 
downtown area they are nearly randomly distributed. This shows for the first 
time that communities within cities are not geographically determined. 

These results are confirmed by the measurement of $\langle r_{com}\rangle$
(average distance between two towers belonging to the same community) and
$\langle r_{rand}\rangle$ (average distance between two random towers), which 
are reported in table \ref{tab:Average-distance-between}. Details on the
calculation of both distances can be found in the {\bf Methods} section.  While 
$\langle r_{rand}\rangle$ is consistently over 4 times larger than $\langle 
r_{com}\rangle$ in the country scale, the two measures become much more similar 
within cities, quantitatively confirming the visual result on Fig.  \ref{Fig6}.

An additional unexpected finding is that some touristic areas break the general 
country-wide trend. A significant part of the French Riviera and the south 
coast of the island of Corsica belong to the Paris community, even if they are 
far away from the capital city. Same thing happens with Ibiza (western most 
Ballearic island) and Madrid.  In Portugal's Algarve (south coast of the 
country) the effect is not so clear, but there is definitely a higher community 
diversity in the area, and it is possible to find towers belonging to both 
Porto and Lisbon communities. Note that this is unlikely to be a touristic 
seasonal effect, because in France and Portugal the most used tower in a 6 
month period is assigned to the user, and in Spain the billing zipcode is used.  
Since both are reasonable proxies for permanent residency, this effect is more 
likely due
to urbanites who retired to the coast, and even become majority in certain
areas, but still keep their social ties back in the large metropolis.

\subsection*{Distance Metrics and Searchability in Urban Networks }
Network searchablity is related to its links density  \cite{watts2002identity, 
    kleinberg2006complex}.  The density of links $P$ as a function of
nodes distance $S$ determines the necessary condition for network searchablity.
This condition is postulated in the group model framework \cite{kleinberg2006complex}, which
generalizes previous results in hierarchies of social networks \cite{watts2002identity} and spatial
lattices \cite{kleinberg2000navigation}. $P(u,v)$ is the probability of link 
existence
between a pair of nodes $(u,v)$ that
are within distance $S(u, v)$, defined as the size of the smallest group
containing both $u$ and $v$. 

Given the distance distribution of the form $P(u,v) \sim S(u,v)^{-\gamma}$ when 
$\gamma < 1$ the social network is not searchable; if $\gamma=1$ the social 
network is always searchable, and if $\gamma > 1$ the network can be 
searchable. 

\begin{figure}[h]
   \centerline{\includegraphics[width=0.85\textwidth]{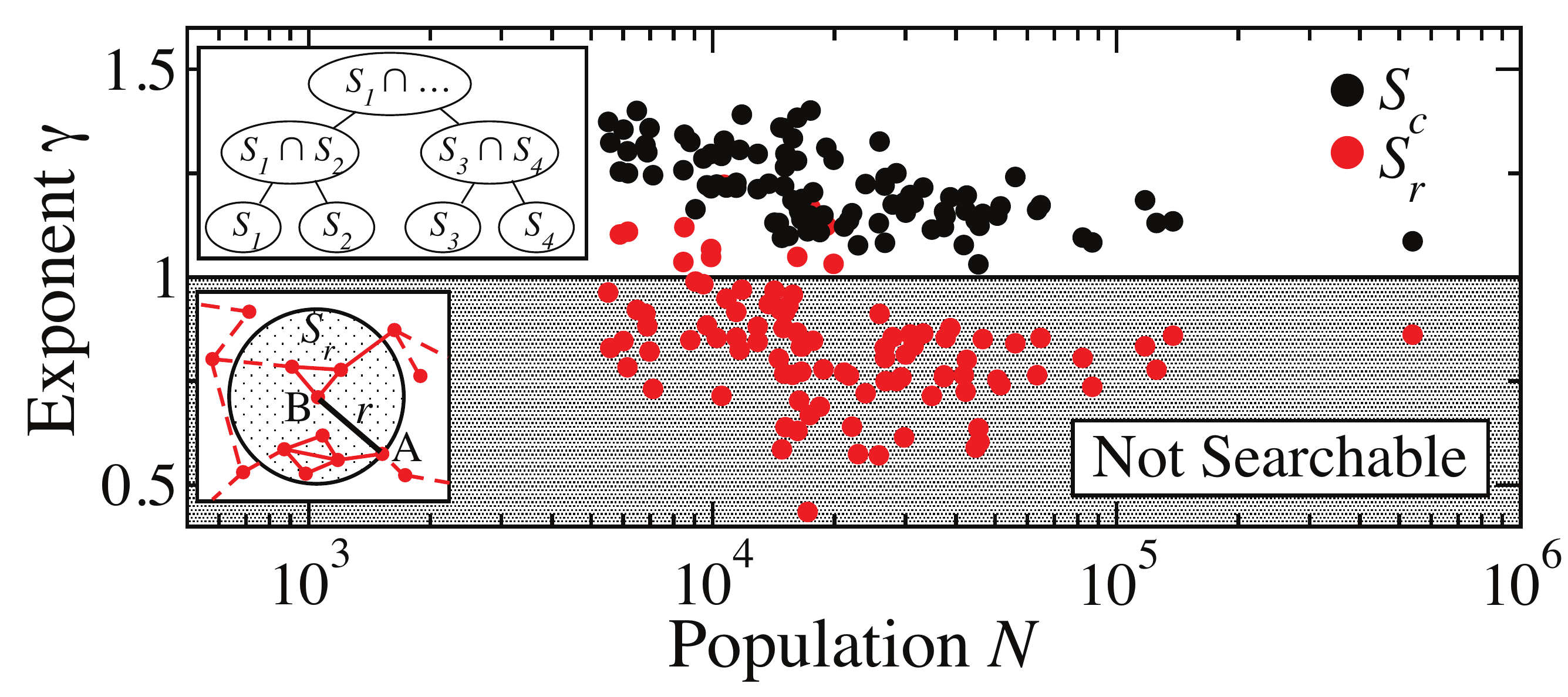}}
   \caption{Comparison of the exponent $\gamma$ for the probability of finding 
       a link between two people as a function of smallest common group size: 
       $p(S_{x}) \sim S_{x}^{-\gamma}$ for 96 cities in France. Groups are 
       constructed either based on geography ($S_{r}$, black) or on community 
       ($S_{c}$, red). }
    \label{Fig7}
\end{figure}

We test this structure in our data using geometric and social distances. We
divide the network into groups of size $S_{X}$ using either geographic balls of
a certain radius $r$ ($X = r$) or existing communities ($X = c$) as illustrated
in the insets of Fig \ref{Fig7}. Then we calculate the probability that two 
nodes that belong to the same group (being that group the smallest they both 
belong to) share a link and how this probability depends on the group size. We 
observe that both functions have the exponent close to $\gamma = 1$, but in the 
groups based on geography these exponents are always below $1$, while the 
exponent is consistently above 1 for communities as shown in Fig.  \ref{Fig7}.  
Although the group-model framework does not capture all of our network 
properties (heterogeneous degree distribution and clustering coefficient) we 
find that our empirical results in urban networks confirm theoretical results 
regarding the conditions for searchability of social networks.

\section*{Discussion}

In summary, we have demonstrated that cities (as defined conventionally by their administrative borders and population size) change the structure of social networks. Interestingly, these findings could be related to urban growth and the economic function of cities \cite{bettencourt2007growth,pan2013urban}.

Taken together, the presented results lead to the following discoveries: (i)
 Communities within cities follow a hierarchical structure that favors social distance over geographic distance.  (ii) While people living within geographic radius including
several cities form a connected network, the same radius within cities leads to
highly 
 clustered components only connected through people in distant parts of the 
 city. This behavior occurs across different cities and regions sizes, 
 highlighting cities as functional entities of the social networks (iii) The
structure of communities (here related to social proximity) and not geographic
distance is what makes social networks searchable within cities. This finding 
is
consistent with experimental results that suggest people do use the 
 profession or name of the target in the final steps to make inferences about 
 his/her education or ethnicity, as a hint to help routing within 
 cities~\cite{dodds2003experimental}.

 This work uncovers an unknown feature of social networks: while at the national level descriptions of social networks  consist of highly connected and 
 geographically close communities, we find that geography plays only a minor 
 role when forming communities within cities. Urban networks consist of geographically dispersed communities. This structure explains why people 
 are able to successfully route in Milgram-like experiments, provided they 
 correctly identify the community of the target.  Our results support the 
 theoretical hypothesis of Kleinberg: the likelihood to find friendships within 
 communities decays as a power-law with increasing community 
 size~\cite{kleinberg2006complex}, confirming that among all possible network 
 configurations, humans have favored those such that a message can reach anyone 
 even if delivered using only local information.  This is a remarkable example 
 of a self-organized structure that allows a small group of individuals to 
 solve a complex problem by cooperating to take advantage of collective 
 knowledge~\cite{moreno2004dynamics, rutherford2013limits}.

\section*{Methods}

\subsection*{Data}
We analyze phone records for a six months period in three countries: France, 
Portugal, and Spain. In total $7$ billion phone interactions are considered. In 
order to build social networks from this data, only links with at least one 
communication per direction are included. This is a common technique in the 
literature \cite{onnela2007structure,onnela2011geographic,lambiotte2008geographical} to avoid both marketing 
callers and misled numbers. The resulting social networks have $18.7$, $1.2$, 
and $5.9$ million users, for  France, Portugal, and Spain respectively. Further 
details are provided in the SI.

\subsection*{Routing Algorithms\label{sec:Algorithms}}

In order to deliver the message, several strategies can be used. In
the following we describe every criteria used in our experiments. 
\begin{description}
\item [{{{RAN}}}] We use random routing as a baseline comparison, by  
    employing depth first search (DFS) into a routing
algorithm, we effectively avoid the message to get into infinite loops.
The application of DFS in the Milgram experiment %way
%it could be implemented in humans 
is quite straightforward: when a participant receives a message, he/she
knows the list of people who already got the message. The participant
will never forward to none of these people, unless all of his/her 
friends
are in the list. In this case, he/she will send the message back to the
person who first sent the message to him. In a tree network, this
would be the case of a branch which has been explored without success
and the search process continues going backwards. Since our social
network is far from being a tree, the number of rolling back events
is low (less than $10^{-6}$ in all of our simulations). 
\item [{{{GEO}}}] This procedure consists of sending the message to
the friend geographically closest to the target. In the intercity
scenario, locations are considered on the municipality level. In the
intracity scenario, tower locations are employed. Note that this discretization
produces a number of ties (two or more friends are at the same distance
from the target). 
\item [{{{DEG}}}] In this case, the message is forwarded to the friend
with the largest number of friends among the candidates. 
\item [{{{COM}}}] In order to mimic social attributes (school, work)
communities are detected in the network. To detect communities in social networks, we use the well-established Louvain method~\cite{blondel2008fast,fortunato2010community,onnela2011geographic, ahn2010link,expert2011uncovering}. This method is a greedy optimization method that attempts to optimize the network modularity by aggregating nodes belonging to the same community and building a new network whose nodes are the communities. This method assigns to each person a set of communities at different hierarchical levels. Although the number of
aggregation levels $L$ depends on the network and it is automatically obtained from the algorithm, in all of our networks
the algorithm provided between 3 and 7 aggregation levels. Note that
this algorithm provides hierarchical communities. If two nodes $i$
and $j$ have a community of level $l$ in common they will share
as well all the communities in higher levels, formally: 
\begin{equation}
i,j\in[1,..N]/c_{il}=c_{jl}\rightarrow c_{ix}=c_{jx}\forall 
x\in[l+1,..,L]
\end{equation}
 where $N$ is the number of people. A person will send the message
to a friend with the lowest possible community level in common with
the target.  While it is arguable that community detection requires global 
information and
such might not be available to participants in a Milgram-like experiment,
recent research \cite{leskovec2012learning} has reported that people are able to 
relate communities detected in their network to certain social attributes and affiliations, thus making communities a reasonable proxy for those unknown attributes in our data set. 

\end{description}

In our experiments, these criteria are combined, by using several
of them to solve ties: this way, we will denote \textit{ran-deg }to
a routing scheme where first the already visited nodes are discarded
from candidates (\emph{ran}), and then those with the highest degree
are chosen (\emph{deg}). If there is still more than one possible
friend after the routing logic is completed, the message is forwarded
to one of these candidates at random. In our \emph{ran-deg} example,
this happens if two or more friends were not previously visited and
have the same degree.

\subsection*{Geographical dispersion of communities}
We found a fundamental difference between the behavior on urban scale and on
the country one: geographical clustering turns out to be more intense
in the intercity scenario than in the intracity one. To reach this conclusion we have calculated the spatial clustering of the communities
by the following steps: 
\begin{itemize}
\item Perform a community detection on the network 
\item Associate the tower to the most common community among that tower's
users.  \item Calculate the average distance $\langle r_{com}\rangle$ between 
any
two towers belonging to the same community 
\begin{equation}
\langle 
r_{com}\rangle=\frac{\sum_{c=1}^{C}\sum_{a=2}^{N_{c}}\sum_{b=1}^{a-1}r(a,b)}{\sum_{c=1}^{C}\sum_{a=2}^{N_{c}}(a-1)}\label{eq:com}
\end{equation}
 where $C$ denotes the number of communities found, $N_{c}$ the
number of towers in community $c$ and $r(a,b)$ the distance between
towers $a$ and $b$. %\item Shuffle the community labels (keeps the community size distribution)
%and repeat the operation to calculate the average distance $\langle d_{r}\rangle$. 

\item Assign communities with the same sizes randomly to the towers and calculate
the average distance (\ref{eq:com}) of the randomized data $\langle 
r_{rand}\rangle$.
\end{itemize}

\begin{acknowledgments}
This work was partially funded by New England UTC Year 23 grant,
the Center for Complex Engineering Systems (CCES) at KACST under the 
co-direction of Anas Alfaris and the MIT-Accenture alliance.  CHY and PJZ 
acknowledge support from Orange Spain (France Telecom Group), PJZ from 
MTM2010-15102 of Ministerio de Ciencia e Innovacion and Q10 0930-144 of the 
UPM, and RMB from Fundaci\'{o}n Caja Madrid (Spain) and project grant 
MINECO-Spain MTM2012-39101.
\end{acknowledgments}

\subsection*{Author contributions}
CHY designed and performed the experiments, analysed data and wrote 
the paper. CMS designed and performed the experiments and wrote the 
paper. TC and ZS performed the initial data extraction. RMB, PJZ and 
MCG designed the experiments, wrote and reviewed the paper. 
 
\subsection*{Competing financial interests}
The authors declare no competing financial interests.

%%%%%%%%%%%%%%%%%%%%%%%%%%%%%%%%%%%%%%%%%%%%%%%%%%%%%%%%%%%%%%%%
%% For figures, put the caption below the illustration.

%
\begin{table}[htp]
\begin{tabular}{lrrrrrrrr}
\hline 
Country  & \% GC  & Nodes $N$  & Links $E$  & $\langle k\rangle$  & $\langle c\rangle$  & $\langle c_{r}\rangle$  & $\langle l\rangle$  & $\langle l_{r}\rangle$\tabularnewline
\hline 
France  & $99.23$ & $18.7\cdot10^{6}$  & $81.3\cdot10^{6}$  & $8.73$ & $0.16$ & $9\cdot10^{-7}$  & $8.52$ & $7.75$\tabularnewline
Portugal  & $96.23$ & $1.21\cdot10^{6}$  & $4.00\cdot10^{6}$  & $6.57$ & $0.26$ & $5\cdot10^{-7}$  & $8.35$ & $7.44$\tabularnewline
Spain  & $95.81$ & $5.92\cdot10^{6}$  & $16.1\cdot10^{6}$  & $5.44$ & $0.21$ & $48\cdot10^{-7}$  & $10.36$ & $9.20$\tabularnewline
\hline 
\end{tabular}\caption{Characteristic properties of the social networks in the studied countries: Size of the giant component (GC), number of users (Nodes)
and relationships (Links), average degree $\langle k\rangle$, average
clustering coefficient $\langle c\rangle$, average shortest path length 
$\langle l\rangle$, and the corresponding
values for random networks with the same size $\langle c_{r}\rangle$
and $\langle l_{r}\rangle$.}
\label{Table:param}
\end{table}

\begin{table}
\begin{centering}
\begin{tabular}{crrr}
\hline 
Network  & $\langle r_{com}\rangle$(km)  & $\langle r_{rand}\rangle$(km)  & $\langle r_{ran}\rangle/\langle r_{com}\rangle$\tabularnewline
\hline 
Portugal  & $64.4$  & $240.1$  & $3.72$\tabularnewline
France  & $115.7$  & $410.71$  & $3.54$\tabularnewline
Spain  & $118.5$  & $521.2$  & $4.39$\tabularnewline
Lisbon (\emph{concelho})  & $3.4$  & $4.31$  & $1.26$\tabularnewline
Paris (\emph{department})  & $4.1$  & $5.7$  & $1.39$\tabularnewline
Madrid (\emph{municipio})  & $3.2$  & $3.46$  & $1.08$\tabularnewline
\hline 
\end{tabular}
\par\end{centering}

\caption{\label{tab:Average-distance-between}Average distance between two
towers belonging to the same community ($\langle r_{com} \rangle$) 
compared to the distance when the communities are randomized ($\langle 
r_{rand} \rangle$).  The geographical effect $\frac{\langle r_{rand} 
    \rangle}{\langle r_{com}\rangle}$
is more pronounced in the nation-wide communities.}
\end{table}

%\clearpage
%% For Tables, put caption above table
%\bibliography{ref1}% Produces the bibliography via BibTeX.

\end{document}